\documentclass{article}
\usepackage[T1]{fontenc}
\usepackage[utf8]{inputenc}
\usepackage{ismir}
\usepackage{amsmath,cite,url}
\usepackage{graphicx}
\usepackage{color}
\usepackage[dvipsnames]{xcolor}
\usepackage{musicography}

\usepackage{multirow}
\usepackage{booktabs}
\usepackage{siunitx}
\newcommand{\upd}[1]{{\color{black}#1}}

\title{Exploring System Adaptations for \\ Minimum Latency Real-Time Piano Transcription}

\multauthor
{Patricia Hu$^1$ \hspace{0.54cm} Silvan David Peter$^1$ \hspace{0.54cm} Jan Schl\"uter$^1$ \hspace{0.54cm} Gerhard Widmer$^{1,2}$}
{
$^1$ Institute of Computational Perception, Johannes Kepler University Linz, Austria\\
$^2$ {LIT AI Lab, Linz Institute of Technology, Austria}\\
{\tt patricia.hu@jku.at}
}

\def\authorname{P. Hu, S. Peter, J. Schl\"uter and G. Widmer}

\begin{document}

\maketitle

\begin{abstract}
Advances in neural network design and the availability of large-scale labeled datasets have driven major improvements in piano transcription.
Existing approaches target either offline applications, with no restrictions on computational demands, or online transcription, with delays of \upd{128}--320\,ms. 
However, most real-time musical applications require latencies below 30\,ms. 
In this work, we investigate whether and how the current state-of-the-art online transcription model can be adapted for real-time piano transcription. 
Specifically, we eliminate all non-causal processing, and reduce computational load through shared computations across core model components and variations in model size. 
Additionally, we explore different pre- and postprocessing strategies, and related label encoding schemes, and discuss their suitability for real-time transcription.
Evaluating the adaptions on the MAESTRO dataset, we find a drop in transcription accuracy due to strictly causal processing as well as a tradeoff between the preprocessing latency and prediction accuracy.
We release our system as a baseline to support researchers in designing models towards minimum latency real-time transcription.

\end{abstract}

\section{Introduction}\label{sec:Intro}
Automatic music transcription (AMT) is the task of transforming audio signals into their symbolic music representation, and is commonly referred to as one of the holy grails in Music Information Retrieval (MIR), given its role in linking the audio and symbolic domain, as well as its relevance to various downstream tasks and musical applications \cite{hawthorne2019enabling, kong2021highresolution}. The transcription of piano solo music is among the most extensively studied tasks, driven by the instrument's well-defined onset characteristics and the availability of large-scale, strongly labeled training data \cite{hawthorne2019enabling}. Consequently, research on automatic piano transcription has seen substantial progress. Notable contributions include \cite{sigtia2016end, kelz2019deep, hawthorne2019enabling, kong2021highresolution}, with the former two setting new benchmarks by leveraging larger architectures, increased model complexity, extended training, and a novel regression-based target encoding approach, leading to higher-resolution piano transcription.

Progress in automatic piano transcription has focused almost exclusively on offline methods, with only a few investigations attempting to solve this task online \cite{kusaka2024mobile, fernandez2023onsets, kwon2024towards}. 
These online systems typically reconfigure offline approaches for block-wise updates while keeping audio representations and network structure, and achieve latencies between \upd{128} and 320\,ms.
Latency values in the hundreds of milliseconds are suitable for musical applications such as subtitling, page turning, or visualizations. 
However, most interactive musical applications require lower latencies, e.g., for digital instruments, a commonly named bound is 10\,ms~\cite{wessel2002problems,DBLP:conf/nime/McPhersonJM16,caspe2025designing}, and for networked ensemble playing latencies of up to 30\,ms are acceptable~\cite{turchet2023relation, lakiotakis2019improving, chew2004musical}. 
A real-time transcription model should enable fluent musical interaction, for which we take 30\,ms as a minimal requirement, and 10\,ms as a goal for imperceptible latency.
Latency stems from various sources: audio buffering, preprocessing, model inference and postprocessing. For an automatic piano transcription system to perform inference in real-time, it must minimize latency in all sources. 
It should adhere to strict causality requirements in both the model architecture and the postprocessing, ensuring that both rely exclusively on past information.

In this work, we adapt a state-of-the-art model for real-time transcription towards minimal latency. 
We do so by allowing only causal processing within the model and reducing computational load by sharing computations across core components of the model.
We further investigate the latencies incurred by widely used pre- and postprocessing strategies, and explore options to mitigate these using a combination of adapted STFT processing, label encoding schemes, loss functions, and causal post-processing.
Our contribution is two-fold: First, we examine the sources of latency and causality violations in the current state-of-the-art system in online piano transcription, and propose and evaluate changes in the modeling and pre- and postprocessing stages for efficient real-time transcription. 
Second, we provide an open-source basis for a low-latency, real-time automatic piano transcription, inviting further research and development in this area.

The remainder of this article is structured as follows: In Section~\ref{sec:Related} we point to related work, one of which will form the our starting point for our adaptations towards minimum latency, which we will therefore focus on in greater detail in Section~\ref{sec:Starting Point}. 
In Section~\ref{sec:Experiments} we present our latency-minimizing adaptations and report the experiments conducted to assess their effect transcription accuracy. 
We discuss the challenges and lessons learned, and an outlook for future work in Section~\ref{sec:Discussion}.

\section{Related work}\label{sec:Related}

As outlined in the previous section, both online and real-time transcription remain largely unexplored. We will discuss three noteworthy contributions \cite{kwon2024towards, fernandez2023onsets, kusaka2024mobile}. 

Fernandez \cite{fernandez2023onsets} proposes a purely convolutional modeling approach that focuses solely on onset and velocity prediction, achieving a latency between 4 and 9 seconds.

Kwon et~al.\ \cite{kwon2024towards, kwon2020polyphonic, jeong2020real} propose an autoregressive neural network for efficient online piano transcription.
The architecture comprises two main components: an acoustic module consisting of a stack of convolutional layers with frequency-conditioned FiLM layers, and a note sequence module consisting of pitchwise LSTMs, a multi-state softmax output (for different note states: onset, sustain, re-onset, offset, and off), and an autoregressive connection from the note state output of the previous time step to the current sequence module input. The authors propose various architectures that balance accuracy and latency. Overall, their models achieve latencies from \upd{128} to 320\,ms.

Kusaka and Maezawa \cite{kusaka2024mobile} introduce Mobile-AMT, a framework designed to tackle both real-time processing and generalization to unseen recording environments in automatic piano transcription. 
They optimize a state-of-the-art offline transcription model \cite{kong2021highresolution} by replacing its conventional convolutional components with lightweight, efficient alternatives \cite{howard2019searching} and train it using a data augmentation scheme that simulates four distinct acoustic distribution shifts. 
The resulting model sets the current state of the art in online automatic piano transcription, achieving F1 scores comparable to offline state-of-the-art models while being robust to in-the-wild recordings. Its latency is reported as 174\,ms -- but an apparent oversight in the architecture increases latency to 10\,s. We use this method as our starting point and detail it in the next section.


\section{Starting Point}\label{sec:Starting Point}

As Mobile-AMT \cite{kusaka2024mobile} represents the current state of the art in real-time piano transcription, we use this method as the foundation and reference method for our adaptations towards minimum-latency transcription. To provide context for these modifications, we first outline the structure of the model, and particularly focus on the modeling aspects that violate causality requirements for real-time systems and necessitate adaptation.

\subsection{Model Architecture}
Mobile-AMT \cite{kusaka2024mobile} is a lightweight adaptation of the state-of-the-art offline piano transcription model by Kong et~al.\ \cite{kong2021highresolution}. It replaces all conventional convolutional blocks with MobileNet \cite{howard2019searching} equivalents, which consist of depthwise separable convolutions that reduce computational complexity while maintaining representational power. Additionally, Mobile-AMT removes all bi-directional flows in recurrent model layers, and drops one of originally four acoustic model core stacks (the one for note offset prediction, which is instead conditioned on the frame and onset output\footnote{The offset prediction is omitted during inference, but no additional details are provided on whether or how the postprocessing is adjusted to account for this missing information.}). All optimization and activation layers are retained from the original offline model. With these modifications, the authors report a resulting latency of 174\,ms, and argue their model to be capable of real-time inference.

Apart from depthwise separable convolutions, Mobile-AMT also adopts MobileNet's Squeeze-and-Excitation (SE) layers to dynamically recalibrate channel-wise features. The squeeze operation involves global pooling over all spatial dimensions, and therefore relies on information from the entire feature map, making the operation non-causal. As Mobile-AMT processes its input in 10-second blocks, the squeeze operation adds 10 seconds of latency, which is not accounted for in the authors' calculations.

\subsection{Postprocessing}\label{subsec:mobileamt_postprocessing}
Mobile-AMT uses the same regression target encoding format as its underlying offline transcription model \cite{kong2021highresolution}, resulting in incrementally increasing and decreasing onset and offset targets over a sequence of frames instead of binary, pointwise classification targets.

While this target encoding format allows for high, sub-frame resolution for onset and offset detection, it necessitates non-causal postprocessing, as the detection of an onset/offset relies on both past and future frames. The authors of Mobile-AMT adopt the same postprocessing strategy as \cite{kong2021highresolution}, and account for it in their overall latency calculation.

\subsection{Training and Evaluation Setup}
Mobile-AMT is trained on 10-second segments of 16\,kHz audio, which are transformed into 229-binned mel spectrograms after an STFT with a 2048-sample Hann window and a hop size of 320 samples. The loss function for the onset, offset and frame targets is binary cross-entropy (BCE), and velocities are trained using mean squared error (MSE). 

Mobile-AMT proposes a data augmentation scheme for training to enhance robustness to real-world, in-the-wild recordings, therefore training duration depends on the data augmentations applied. 
For the non-augmented baseline, Mobile-AMT is trained for 3000 epochs on the MAESTRO dataset \cite{hawthorne2019enabling} with a batch size of 16 and uses the Adam optimizer with a learning rate of 0.001 annealed to 0 following a cosine schedule \cite{lakiotakis2019improving}. 




\section{Real-Time Adaptations}\label{sec:Experiments}

As our main contribution, we implement and evaluate adaptations of the starting point that aim to reduce the system's latency. 
We form three groups of experiments: adapting the training and postprocessing, adapting the audio preprocessing, and adapting the network architecture. 
We conduct each experiment with a reduced training time of 500 epochs (30k updates, approximately one-sixth of the total training time reported in our reference method \cite{kusaka2024mobile}) to save computational resources, as most effects become evident already during the early training stages. 
We perform one final comparison of a model in which we combine selected modifications, and compare this to our reference method, with both models trained for 2000 epochs.

For all our experiments, unless otherwise noted, we train on 3-second segments of audio at 16\,kHz, transformed into 229-bin mel spectrograms after an STFT with a Hann window of 2048 and a hop size of 160 samples. We double the frame rate compared to Mobile-AMT to allow for more frequent updates, and therefore reduced latency.

Apart from the training duration and data encoding, we follow the same training scheme as used in the non-augmented setup in our reference method, i.e., all adapted models are trained and evaluated on the MAESTRO v3.0 dataset using \textit{mir\_eval}\cite{raffel2014mir_eval}.
As our work focuses on minimizing delay in real-time transcription, we mostly focus on the note onset metrics. Except for the final comparison, we only evaluate on the validation set. 
\upd{
Since our goal is to achieve minimal latency for real-time musical interactions, we 
assess inference performance on stricter timing tolerances (10--30\,ms) to reduce the mismatch between system evaluation and target performance.
}

\subsection{Training and Postprocessing}\label{sec:TP-exp}

In Mobile-AMT, the training target for an onset or offset extends over multiple time steps, forming a triangle centered on the target annotation. Originally proposed by Kong et~al. \cite{kong2021highresolution}, such triangles allow expressing positions at a higher resolution than the frame rate. However, predicting such triangles requires lookahead in the model, as the output must increase before the actual event. Furthermore, interpreting the predictions requires lookahead in postprocessing in order to find each triangle's maximum. As a preparation for switching to a causal version of Mobile-AMT with causal postprocessing, we thus replace the triangular targets with binary targets that are active only at the frame that is closest to an annotation. 

As the changed label encoding results in a heavily imbalanced binary classification task, particularly for onset and offset targets, we try weighting their positive occurrences with a factor of 10 in the binary cross-entropy loss \upd{\cite{bittner2022lightweight}}. Additionally, pointwise classification targets penalize small temporal deviations more strongly than triangular targets. To account for this, we test applying a shift-tolerant loss recently proposed for beat detection \cite{foscarin2024beat}, with a tolerance of $\pm$1 frame ($\pm$ 10\,ms at our frame rate).

We also modify the postprocessing to ensure that the prediction for the current frame relies only on past information.
First, we binarize the output probabilities: for onset and offset targets, an activation is detected when the current frame exceeds a given threshold while the previous frame does not. A frame activation is recorded if the current frame surpasses the threshold. Next, we eliminate re-onsets of the same pitch that occur within a predefined minimum re-onset distance. Finally, we determine the offset of an active note based on the earlier occurring one of either frame inactivity or offset activity.

\begin{table}[t!]
    \centering
    \small
     \resizebox{\columnwidth}{!}{
    \renewcommand{\arraystretch}{1}
    \begin{tabular}{l
  r@{~}r
  r@{~}r
  r@{~}r}
\toprule
\textbf{Tolerance} & \multicolumn{2}{c}{\textbf{10\,ms}} & \multicolumn{2}{c}{\textbf{20\,ms}} & \multicolumn{2}{c}{\textbf{30\,ms}} \\
\midrule
\texttt{Exp.} & \multicolumn{6}{l}{Postprocessing onset threshold: 0.45} \\
\midrule

\texttt{TP1} & \hphantom{0}9.58 & $\pm$2.29 & 25.80 & $\pm$~\hphantom{0}4.83 & 47.44 & $\pm$~\hphantom{0}7.30 \\
\texttt{TP2} & 29.84 & $\pm$8.37 & 47.64 & $\pm$~11.76 & 50.28 & $\pm$~11.83 \\
\texttt{TP3} & 13.95 & $\pm$4.91 & 34.34 & $\pm$~\hphantom{0}8.05 & 45.10 & $\pm$~\hphantom{0}8.33 \\
\texttt{TP4} & 27.24 & $\pm$7.47 & 55.49 & $\pm$~\hphantom{0}9.73 & 64.95 & $\pm$~10.06 \\
\texttt{TP5} & 16.00 & $\pm$5.32 & 41.02 & $\pm$~\hphantom{0}8.50 & 55.40 & $\pm$~\hphantom{0}8.53 \\

\midrule

\texttt{Exp.} & \multicolumn{6}{l}{Postprocessing onset threshold: 0.55} \\
\midrule

\texttt{TP1} & \hphantom{0}13.42 & $\pm$2.60 & 35.07 & $\pm$~\hphantom{0}5.31 & 56.02 & $\pm$~\hphantom{0}8.14 \\
\texttt{TP2} & \hphantom{0}23.56 & $\pm$8.41 & 35.50 & $\pm$~11.44 & 36.87 & $\pm$~11.49 \\
\texttt{TP3} & \hphantom{0}17.04 & $\pm$5.52 & 40.54 & $\pm$~\hphantom{0}8.26 & 51.46 & $\pm$~\hphantom{0}8.29 \\
\texttt{TP4} & \hphantom{0}27.42 & $\pm$7.52 & 54.95 & $\pm$~10.18 & 63.67 & $\pm$~10.72 \\
\texttt{TP5} & \hphantom{0}17.46 & $\pm$5.59 & 44.19 & $\pm$~\hphantom{0}8.53 & 58.68 & $\pm$~\hphantom{0}8.61 \\

\midrule
\texttt{Exp.} & \multicolumn{6}{l}{Postprocessing onset threshold: 0.65} \\
\midrule
\texttt{TP1} & \hphantom{0}19.10 & $\pm$3.12 & 43.88 & $\pm$~\hphantom{0}7.02 & 59.93 & $\pm$~\hphantom{0}9.77 \\
\texttt{TP2} & \hphantom{0}14.32 & $\pm$6.70 & 20.52 & $\pm$~\hphantom{0}8.68 & 21.07 & $\pm$~\hphantom{0}8.70 \\
\texttt{TP3} & \hphantom{0}20.70 & $\pm$5.90 & 46.48 & $\pm$~\hphantom{0}8.28 & 56.72 & $\pm$~\hphantom{0}8.28 \\
\texttt{TP4} & \hphantom{0}27.18 & $\pm$7.65 & 53.65 & $\pm$~10.71 & 61.55 & $\pm$~11.36 \\
\texttt{TP5} & \hphantom{0}18.74 & $\pm$5.80 & 46.80 & $\pm$~\hphantom{0}8.60 & 61.05 & $\pm$~\hphantom{0}8.89 \\

\bottomrule
\end{tabular}

    }
    \caption{Comparison of note onset F1 scores (mean $\pm$ std each over three experimental runs) on the MAESTRO v.3 validation set across \upd{three} onset tolerances for different target encodings, (weighted) loss functions and onset thresholds in postprocessing.}
    \label{tab:TL_results}
\end{table}

We test the change in label encoding in combination with different (weighted) loss functions in five different experimental setups: In \texttt{TP1} we train our reference model with the original loss functions and regression target encoding scheme as proposed by the authors \cite{kusaka2024mobile}. In \texttt{TP2} we use classification targets, and in \texttt{TP3} we additionally apply a weight factor $10$ on both the onset and offset targets. \upd{Finally,} we apply the shift-tolerant BCE loss (with a tolerance of $\pm$10\,ms), \upd{either unweighted (\texttt{TP4}) or weighted again by a factor of $10$ (\texttt{TP5})}. All five setups use our strictly causal postprocessing described above. The model architecture remains as proposed in Mobile-AMT.

Table~\ref{tab:TL_results} lists the note onset F1 scores on the MAESTRO v3.0 validation set for the different training targets and loss functions, combined with different onset threshold values applied during postprocessing, as evaluated on different onset tolerance thresholds. \upd{We choose to evaluate at lower tolerance thresholds as they reflect a real-time system's practical responsiveness better than the commonly used $\pm$~50ms do, which might mask latency issues by crediting the model for detections that would feel delayed to a user in an interactive scenario.} 
\upd{Learning pointwise binary targets without any further modification (\texttt{TP2}) proves superior to short sequential regression (\texttt{TP1}), provided a low enough onset threshold during postprocessing. The effect of lower onset detection thresholds is somewhat mitigated by weighting positive labeled examples higher (\texttt{TP3}), and virtually fully eliminated by using a shift-tolerant loss (\texttt{TP4}). There is no further improvement when combining weights and shift-tolerance (\texttt{T5}).}

For the next experiments, we use binary targets with weights. While the shift-tolerant loss performed favorably here, our goal is to use a causal model, for which shift tolerance could result in systematically delayed predictions. 

\begin{figure}[t]
\includegraphics[width=\linewidth]{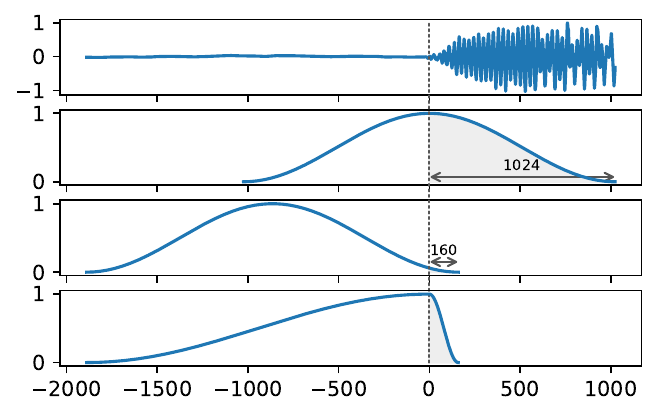}
\caption{Mobile-AMT centers STFT windows on the time point to predict for, incurring a delay of 1024 samples. We shift the window to reduce the delay to 160 samples, and change the window function to better use that limited amount of future information. \label{fig:windows}}
\end{figure}

\subsection{Audio Preprocessing}\label{sec:W-exp}

Mobile-AMT processes audio in spectrogram frames of 2048-sample windows centered on the time points to predict events for. 
Thus, even with a causal model that does not process information from future frames, in real-time inference, every time a complete audio buffer of 2048 samples is filled, inference will be triggered to predict events that are already 1024 samples (64\,ms) in the past.
Figure~\ref{fig:windows} illustrates this: the top row shows an audio waveform, the second row a typical STFT window centered at the onset.
Shorter filters reduce this delay, as the delay is fixed to half the filter length, but this comes at the cost of lower frequency resolution, which we should avoid:
With a 2048-sample STFT at 16\,kHz, the bin width is 7.8\,Hz, which is already too coarse to achieve semitone precision for the lowest piano notes (A0 at 27.5\,Hz, B\musFlat{}0 at 29.14\,Hz).

We can however reduce this delay to a lower number $n_{s}$ of samples (e.g., 160 samples or 10\,ms) while keeping the window length and frequency resolution unchanged, by shifting windows so they end $n_{s}$ samples after their reference point instead of being centered.
The third row in Figure~\ref{fig:windows} shows the Hann window shifted to $n_s=160$ samples. The Hann window strongly attenuates the boundaries, the right one of which now contains highly relevant information for the prediction.
To mitigate this unwanted attenuation, we can replace the Hann window with an asymmetric window that tapers $(2048-n_{s})$ samples before and $n_{s}$ samples after the reference point.
The last row in Figure~\ref{fig:windows} illustrates this windowing function for a 1888/160 sample asymmetry.
Note how we keep more information from the incoming samples in the gray shaded area under the window function\upd{, albeit at the cost of increasing spectral leakage (by about 20\,dB for $n_s=160$)}.

To experiment with different windowing configurations for reducing the delay in audio preprocessing,
we modify our reference method to apply only causal processing, as allowing the model access to future frames would render our interventions meaningless. 
Specifically, we make each convolution causal, so the model’s receptive field of 9 frames extends 8 frames into the past, rather than splitting 4 frames into the past and 4 frames into the future. 
Additionally, we remove the Squeeze-Excitation layers of the MobileNet V3 blocks, which perform global average pooling over both past and future frames in an excerpt.

\begin{table}[btp!]
    \centering
    \small
     \resizebox{\columnwidth}{!}{
    \renewcommand{\arraystretch}{1}
\begin{tabular}{l@{~~}l@{~}lr@{~\hphantom{1}}rr@{~\hphantom{1}}rr@{~\hphantom{1}}r}

    \toprule
    \multicolumn{3}{c}{\textbf{Tolerance}} & \multicolumn{2}{c}{\textbf{10\,ms}} & \multicolumn{2}{c}{\textbf{20\,ms}} & \multicolumn{2}{c}{\textbf{30\,ms}}\\
    \midrule
    
    \texttt{H1} & Hann & 64\,ms & 17.43 & $\pm$~5.10  & 34.65 & $\pm$~7.40  & 39.86 & $\pm$~7.45  \\
    \texttt{H2} & Hann & 10\,ms & 0.00 & $\pm$~0.01  & 0.00 & $\pm$~0.02  & 0.04 & $\pm$~0.10  \\
    \texttt{T1} & asym. & 10\,ms & 22.25 & $\pm$~5.17  & 25.21 &  $\pm$~5.20  & 25.84 & $\pm$~5.18 \\
    \texttt{T2} & asym. & 20\,ms & 28.61 & $\pm$~7.09  & 33.76 & $\pm$~7.01  & 34.65& $\pm$~6.89  \\
    \texttt{T3} & asym. & 30\,ms & 27.61 & $\pm$~6.79 & 37.91 & $\pm$~7.54  & 39.43 & $\pm$~7.33  \\
    \texttt{T4} & asym. & 40\,ms & 24.39 & $\pm$~6.50  & 37.51 & $\pm$~7.81  & 39.87 & $\pm$~7.60  \\
    \texttt{T5} & asym. & 50\,ms & 20.99 & $\pm$~6.02  & 36.88  & $\pm$~7.75  & 40.47 & $\pm$~7.59  \\
    \texttt{ST} & asym. & 10\,ms & 0.41 & $\pm$~0.40  & 1.57 & $\pm$~0.82   & 11.86 & $\pm$~4.25  \\
    \bottomrule
\end{tabular}

    }
    \caption{Note onset F1 scores on the MAESTRO v.3 validation set for different windowing functions. The last row additionally uses a shift-tolerant training loss.}
    \label{tab:W_results}
\end{table}

\begin{figure}[btp]
\includegraphics[width=\linewidth,trim=0 0 0 20,clip]{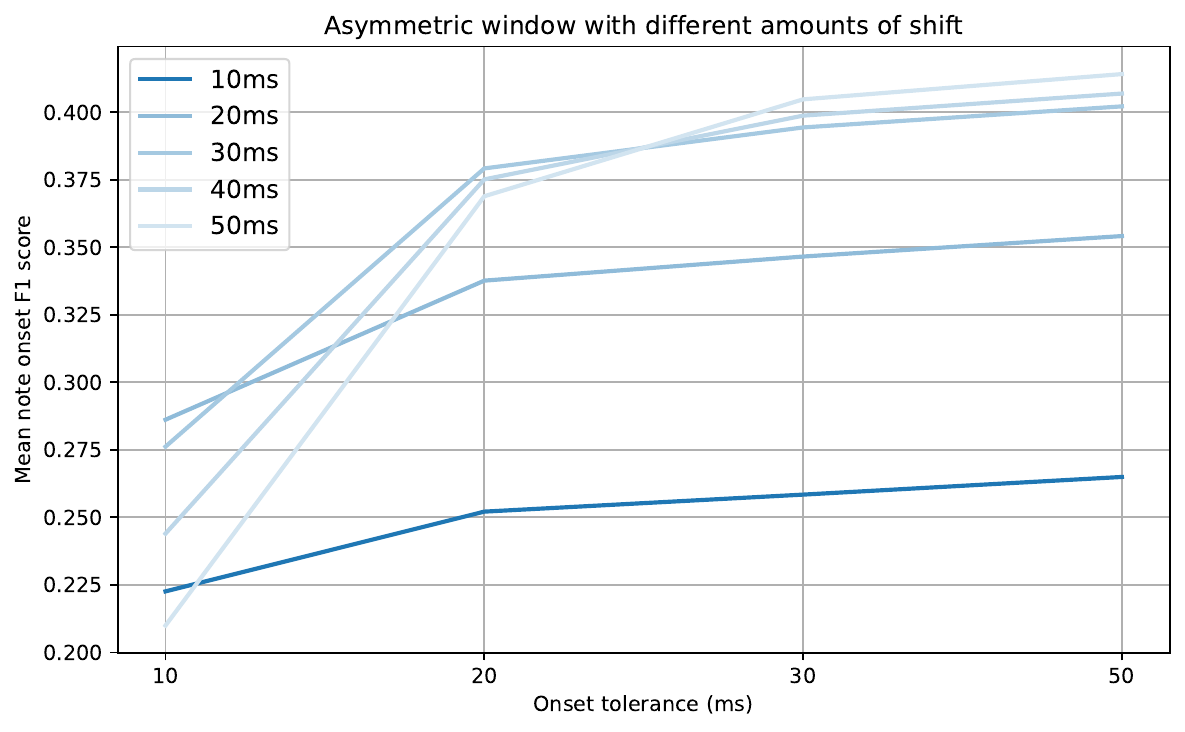}
\caption{Note onset F1 scores (means only) for different window delays and onset tolerance thresholds.
\label{fig:W_results}}
\end{figure}

Table~\ref{tab:W_results} shows our results.
\upd{
The original centered Hann window with our causal model (\texttt{H1}) performs worse than our non-causal starting point (\texttt{TP3} in Table~\ref{tab:TL_results}). Shifting the Hann window from a delay of 64\,ms to a delay of 10\,ms (\texttt{H2}) seems to completely attenuate usable information in the frames.
}
Using an asymmetric window (\texttt{T1}) improves performance, but still falls behind the centered Hann window. Successively increasing the delay up to 50\,ms, we see a strong improvement (\texttt{T2}--\texttt{T5} and Figure~\ref{fig:W_results}). For a delay of 30\,ms or more, we match performance of the centered Hann window at an evaluation onset tolerance of \upd{30\,ms}. For stricter tolerances, shifted asymmetric windows of 20\,ms delay or more surpass the centered Hann window.

We also take the chance to investigate how a shift-tolerant loss of $\pm$1 frame affects results for the causal model. The loss could allow the model to systematically predict events one frame (10\,ms) later than annotated. Surprisingly, using an asymmetric window with 10\,ms of delay, we find that the shift-tolerant loss (\texttt{ST}) performs on par with 30\,ms delay (\texttt{T3}) when admitting an evaluation tolerance of 50\,ms \upd{(not shown in table)}, but breaks down with any stricter tolerance \upd{(as seen in the table)}.



For the third group of experiments, we keep the strictest setting with asymmetric windows at a delay of 10\,ms.

\begin{table}[!ht]
    \centering
    \small
     \resizebox{\columnwidth}{!}{
    \renewcommand{\arraystretch}{1}
\begin{tabular}{lcccc}
    \toprule
    \textbf{Tolerance} & \textbf{10\,ms} & \textbf{20\,ms} & \textbf{30\,ms} \\
    \midrule
    \multicolumn{4}{c}{Note onset F1 mean $\pm$~std} \\
    \midrule
    \texttt{A1} & 20.23 $\pm$~4.29 & 23.03 $\pm$~4.25 & 23.75 $\pm$~4.24 \\
    \texttt{A2} & 25.77 $\pm$~4.65 & 30.66 $\pm$~4.98 & 31.79 $\pm$~4.93 \\
    \texttt{A3} & 21.59 $\pm$~4.79 & 24.48 $\pm$~4.83 & 25.14 $\pm$~4.80 \\
    \texttt{A4} & 21.53 $\pm$~4.31 & 24.51 $\pm$~4.24 & 25.22 $\pm$~4.19 \\
    \texttt{A5} & 19.44 $\pm$~4.20 & 22.39 $\pm$~4.34 & 23.12 $\pm$~4.30 \\
    \texttt{A6} & 25.82 $\pm$~4.48 & 30.52 $\pm$~4.51 & 31.56 $\pm$~4.44 \\ 

    \midrule
    \multicolumn{4}{c}{Note onset and offset F1 mean $\pm$~std} \\
    \midrule
    \texttt{A1} & 3.25 $\pm$~1.31 & 5.13 $\pm$~2.54 & 7.20 $\pm$~3.51 \\
    \texttt{A2} & 3.56 $\pm$~1.29 & 6.17 $\pm$~2.80 & 8.39 $\pm$~4.01 \\
    \texttt{A3} & 5.84 $\pm$~1.41 & 5.84 $\pm$~2.48 & 7.71 $\pm$~3.43 \\
    \texttt{A4} & 5.94 $\pm$~1.31 & 5.94 $\pm$~2.48 & 7.89 $\pm$~3.49 \\
    \texttt{A5} & 4.84 $\pm$~1.11 & 4.84 $\pm$~2.36 & 6.61 $\pm$~3.32 \\
    \texttt{A6} & 6.79 $\pm$~1.15 & 6.79 $\pm$~2.53 & 8.89 $\pm$~3.59 \\ 
    \bottomrule
\end{tabular}

    }
    \caption{Note onset and onset-and-offset F1 scores on the MAESTRO v.3 validation set across \upd{three} onset tolerances for architectural modifications and input representations.}
    \label{tab:A_results}
\end{table}

\subsection{Model Architecture}\label{sec:A-exp}

\upd{In our final group of experiments, we investigate architectural modifications to our reference model.} 
The architecture of Mobile-AMT consists of three acoustic stacks, each consisting of recurrent convolutional blocks. Each stack learns a (onset, frame or velocity) target.
For some targets, the outputs of multiple stacks are concatenated to condition the final predictions.
Compared to their reference offline model \cite{kong2021highresolution}, Mobile-AMT omits the acoustic stack for the offset target\upd{, reusing the stack of the frame target}.

Our experiments involve the following adaptations, each tested independently: First, in \texttt{A1} we (re)introduce a separate offset acoustic stack to explore whether and how it improves offset label prediction. In \texttt{A2} we remove the velocity conditioning on the onsets. Next, we examine whether further streamlining the architecture by sharing \upd{a fourth (\texttt{A3}), half (\texttt{A4}) or all (\texttt{A5}) of the convolutional blocks in}
the model's acoustic stacks affects performance. Lastly, in \texttt{A6} we examine the effect of 
training on the original 10 seconds sequence length.

Table~\ref{tab:A_results} presents the note onset and note onset-and-offset F1 scores of the model and data adaptations on the MAESTRO validation set. Overall, it is evident that the combined impact of binary, heavily imbalanced pointwise targets, causal modeling, and shifted asymmetric window results in a significantly harder learning problem, with the same training duration (500 epochs) leading to significantly poorer scores than the base case (\texttt{TP1} in Table~\ref{tab:TL_results}). However, across all experimental setups in Section \ref{sec:TP-exp} compared to the current one, all our causal modifications demonstrate significantly stronger robustness to decreasing tolerance thresholds, \upd{which is important to guarantee low latency in predictions,} and therefore appear promising for further training.

Furthermore, when comparing all model architecture modifications (\texttt{A1-5}) on note onset and note-onset-and-offset F1 score, we observe two unexpected model behaviours: First, adding a separate offset stack (\texttt{A1}) does not improve offset prediction.
As our postprocessing detects a note offset as the earlier of either offset activation or frame inactivation, we hypothesize that frame activity is sufficiently learned 
to compensate for the absence of an offset acoustic stack. Second, removing the velocity conditioning on onset prediction (\texttt{A2}) results in a strong improvement in onset prediction. 
\upd{Furthermore}, sharing the acoustic stack across increasing proportions (\texttt{A3-5}) does not appear to hinder the model's ability to learn meaningful representations. 
\upd{Finally, experiment \texttt{A6} suggests that the model benefits from the larger contextual window.
}


\subsection{Final comparison}\label{sec:F-exp}

For our final comparison, we proceed with the following data and model configurations: we continue with the (160 samples) shifted asymmetric window for the STFT (\texttt{T1} in Sec. \ref{sec:W-exp}), remove the velocity conditioning (\texttt{A2}) and share all convolutional layers in the acoustic stack across all targets (\texttt{A5}). \upd{Mobile-AMT uses the original non-causal postprocessing described in Section~\ref{subsec:mobileamt_postprocessing}, while our model use the causal postprocessing introduced in Section~\ref{sec:TP-exp}.}

\begin{table*}[ht!]
    \centering
    \small
\begin{tabular}{lc r@{~}r r@{~}r r@{~}r r@{~}r r@{~}r r@{~}r}
\toprule
\multicolumn{2}{c}{} & \multicolumn{6}{c}{\textbf{Note Onset}} & \multicolumn{6}{c}{\textbf{Note Onset with Offset}} \\  
\textbf{Model} & \textbf{Tol. (ms)} 
& \multicolumn{2}{c}{\textbf{Precision}} 
& \multicolumn{2}{c}{\textbf{Recall}} 
& \multicolumn{2}{c}{\textbf{F1}} 
& \multicolumn{2}{c}{\textbf{Precision}} 
& \multicolumn{2}{c}{\textbf{Recall}} 
& \multicolumn{2}{c}{\textbf{F1}} \\
\midrule

Causal-AMT & 10 & 43.51 & $\pm$ 6.87  & 25.60 & $\pm$~\hphantom{0}8.64  & 31.55 & $\pm$ 7.76  & \hphantom{0}6.86 & $\pm$ 2.15  & 4.11 & $\pm$ 2.03  & 5.03 & $\pm$ 2.08 \\

Mobile-AMT & 10 & 22.70 & $\pm$ 5.82  & 15.57 & $\pm$~\hphantom{0}2.96  & 18.26 & $\pm$ 3.51  & \hphantom{0}3.27 & $\pm$ 1.19  & 2.32 & $\pm$ 0.94  & 2.69 & $\pm$ 1.00 \\

Causal-AMT & 20 & 50.86 & $\pm$ 7.52  & 29.71 & $\pm$~\hphantom{0}9.11  & 36.70 & $\pm$ 7.96  & \hphantom{0}10.89 & $\pm$ 3.91  & 6.34 & $\pm$ 2.90  & 7.86 & $\pm$ 3.16 \\

Mobile-AMT & 20 & 59.28 & $\pm$ 7.41  & 41.87 & $\pm$~\hphantom{0}8.44  & 48.52 & $\pm$ 6.96  & \hphantom{0}11.03 & $\pm$ 4.15  & 7.98 & $\pm$ 3.73  & 9.16 & $\pm$ 3.84 \\

Causal-AMT & 30 & 51.85 & $\pm$ 7.49  & 30.24 & $\pm$~\hphantom{0}9.07  & 37.38 & $\pm$ 7.85  & \hphantom{0}14.49 & $\pm$ 5.72  & 8.33 & $\pm$ 3.93  & 10.37 & $\pm$ 4.42 \\

Mobile-AMT & 30 & 81.17 & $\pm$ 6.29  & 57.84 & $\pm$~12.79 & 66.80 & $\pm$ 9.78  & \hphantom{0}18.42 & $\pm$ 6.89  & 13.27 & $\pm$ 6.09  & 15.26 & $\pm$ 6.30 \\

\bottomrule
\end{tabular}

    \caption{Final comparison between our implementation of Mobile-AMT and our modified minimal-latency, strictly causally adapated model.}
    \label{tab:LR_final}
\end{table*}

Table~\ref{tab:LR_final} summarizes the results over different onset (and offset) thresholds for note onset and onset-and-offset metrics. As expected, Mobile-AMT outperforms our modified causal model across all metrics, with a significant margin. Upon reviewing all experiments conducted, we conclude that the largest performance drops are attributed to the shifted window function and the causal convolutions in our model. When comparing Mobile-AMT and our adapted model across various onset tolerance thresholds, we observe, similar to the previous experiment, that while our modified causal model predicts fewer targets with lower accuracy overall, it demonstrates higher precision and robustness when evaluated at stricter timing tolerances.




\section{Discussion and Outlook}\label{sec:Discussion}

In this work, we investigate whether and how the current state of the art in real-time piano transcription can be adapted to achieve minimum-latency automatic piano transcription suitable for real-time musical interaction. 


What latency is suitable cannot be answered universally\upd{, so our choice of 10--30\,ms is worthy of discussion.
While 10\,ms is suggested in digital instrument design~\cite{wessel2002problems,DBLP:conf/nime/McPhersonJM16,caspe2025designing},} thresholds for latency perception vary depending on the musical situation, task, and instrument:
for percussive digital instruments, decreased ratings for values of 20\,ms and above were found~\cite{jack2018custominstrument}, instrument-specific thresholds between below 10\,ms and 40\,ms are reported in a live monitoring setting~\cite{lester2007the}, and about 30\,ms were found for gestural control~\cite{DBLP:conf/icmc/Maki-PatolaH04}.
Offsets as low as 6\,ms may be perceived in simple isochronously spaced stimuli~\cite{friberg1995time}, while other researchers found just noticeable latency differences at 27\,ms and higher~\cite{DBLP:conf/audio/0008ABW24}.
%
Transcription-enabled real-time applications like interactive accompaniment or generative improvisation are more akin to ensemble playing than direct instrument control.
In networked musical contexts, researchers typically aim for 20--30\,ms of latency to meet performance conditions that mirror traditional in-person ensembles \cite{turchet2023relation, lakiotakis2019improving}. 
However, studies also found that musicians may be able to compensate for latencies up to 50\,ms~\cite{chew2004musical, dahl2001player} or even 100\,ms~\cite{10.1145/982484.982506} for one piano piece, a value that was deemed ``neither musical nor interactive'' in another study~\cite{10.1525/mp.2006.24.1.49}.
A real-time transcription model should not only be tolerable but enable fluent musical interaction, \upd{so we took} 30\,ms as a minimal requirement, and 10\,ms as a goal for imperceptible latency.

We investigate multiple adaptations to reduce latency, including label encoding with causal postprocessing, \upd{shifted asymmetric window functions} during preprocessing, and architectural modifications that enforce causal processing within the model. Additionally, we reduce the model size 
\upd{(from~320 to~160 GFLOPs for 3\,seconds of input)}
by sharing computations across core model components for all targets.

In a first set of experiments, we assess the impact of regression versus classification loss encodings for non-causal models.
The original regression targets only make sense in conjunction with a lookahead as the targets begin to increase several frames before the actual onset which is impossible for a causal model to predict.
To mitigate the cost in training stability and accuracy incurred by localized, causal-ready targets, we experiment with loss functions that weight the active frames over the inactive frame to combat label imbalance, and loss functions that are tolerant to small temporal shifts.
We find that the weighted classification losses approximate the baseline, and the shift-tolerant losses reach the same level in the absence of targets requiring lookahead.

In a second experiment, we investigate the delay incurred by the computation of audio feature representations.
Specifically, we look at STFT windows and their corresponding centered targets.
Transcription requires high frequency resolution for pitch estimation which requires large windows.
Centering the targets results in an often overlooked delay of half the window length, 64\,ms in our case.
We test configurations of shifted windows along with asymmetric windowing functions that do not attenuate the most recent samples.
We find that aggressively shifted windows at 10\,ms do deteriorate the transcription accuracy by a lot, yet at 30\,ms, we reach comparable performance to an unshifted causal model.
Here, a shift-tolerant loss does not improve performance.
At the same time, configuring the model architecture for strictly causal processing also deteriorates performance with respect to the baseline with more than 100\,ms of lookahead.

In a third experiment, we assess different model architectures and their impact.
We observe that sharing the convolutional components of the acoustic stack across different target types proves beneficial. 
We hypothesize that the local acoustic features captured in the convolutional layers of the acoustic model can be effectively learned independently of sequential information, making them invariant to the target type. 
Furthermore, removing the velocity conditioning on the onsets strongly improves the accuracy of onset predictions.

Overall, we find that we can compensate well for algorithmic issues: we can scale the model and use lookahead-free targets without a major drop in performance.
What proves difficult, however, is to render the model strictly causal and to effectively process the incoming audio without loss of relevant information.
For a latency of 10\,ms, it would be required that the model predicts pitches with at most 10\,ms of incoming audio samples.
For onsets of the lowest two octaves on the piano, this means that there is not even a full period of the fundamental frequency present in the samples\upd{, and predictions may need to rely on harmonic partials}. 
Along with the transient phase and the consequently blurry STFT frame, this leads to an increasingly hard transcription task.
We hope that these findings and pinpointed challenges will contribute to future research on real-time, minimum latency automatic piano transcription.
    
\upd{While this study primarily focuses on the algorithmic performance and robustness of a real-time transcription model, we acknowledge that a detailed analysis of processing time—including both network inference and preprocessing—across different hardware platforms remains an important area for future work to allow for the practical deployment of a real-time transcription system in real-world scenarios.
Likewise, we want to take a closer inspection into the design of the underlying window function and filter bank, in order to find an appropriate balance between reducing prediction delay and increasing future context, all while maintaining the desired STFT properties. 
Lastly, across all experimental groups, our system adaptations consistently outperformed the baseline at lower timing tolerance, which we consider a desirable property worthy of further investigation.
}

\section{Acknowledgements}
This research acknowledges support by the European Research Council (ERC), under the European Union's Horizon 2020 research and innovation programme, grant agreement No.~101019375 \textit{Whither Music?}. The LIT AI Lab is supported by the Federal State of Upper Austria.

\bibliography{ISMIRtemplate}

%
%
%
%

\end{document}